\documentclass{aa}  
\usepackage{hyperref}
\hypersetup{
  colorlinks   = true, %Colours links instead of ugly boxes
  urlcolor     = blue, %Colour for external hyperlinks
  linkcolor    = blue, %Colour of internal links
  citecolor   = blue %Colour of citations
}
\usepackage{graphicx}
\usepackage{multirow}
\usepackage{longtable}
\usepackage{threeparttable}
\usepackage{hyperref}

\hypersetup{
    colorlinks=true,
    linkcolor=blue,
    filecolor=magenta,
    urlcolor=blue,
    citecolor=blue,
}
\usepackage{txfonts}
\begin{document} 

    \title{Disentangling the association of PAH molecules \\ with star formation}%Insights from JWST and UVIT}

    \subtitle{Insights from JWST and UVIT}
    
    \titlerunning{Association of PAH molecules with star formation}

    \authorrunning{Ujjwal et al.}
%\title{ How star formation affects the PAHs in the nearby Universe}

%   \subtitle{}

   \author{K. Ujjwal  \inst{1}\thanks{ujjwal.krishnan@res.christuniversity.in},  
   Sreeja S Kartha\inst{1},
   Akhil Krishna R\inst{1},
   Blesson Mathew\inst{1},
   Smitha Subramanian\inst{2},
   Sudheesh T P\inst{1}, \\ \and Robin Thomas\inst{1}\\}

   \institute{Department of Physics and Electronics, CHRIST (Deemed to be University), Bangalore 560029, India\\
\and Indian Institute of Astrophysics, Sarjapur Road, Koramangala, Bangalore 560034, India
             }

   \date{Received xxxx; accepted xxxx}

% \abstract{}{}{}{}{} 
% 5 {} token are mandatory
 
  \abstract
   {Polycyclic Aromatic Hydrocarbons (PAHs) are ubiquitous complex molecules in the interstellar medium and are used as an indirect indicator of star-formation. On the other hand the ultraviolet (UV) emission from the young massive stars directly traces the star formation activity in a galaxy. The James Webb Space Telescope (JWST), along with the UltraViolet Imaging Telescope (UVIT), opened up a new window of opportunity to make a better understanding of the properties of the PAH molecules associated with the star-forming regions.}   
   { In this study, we investigate how the resolved scale properties of PAH molecules in nearby galaxies are affected by star-formation.}
   {We analyze the PAH features observed at 3.3, 7.7, and 11.3 $\mu m$ using F335M, F770W, and F1130W images obtained from JWST. These images help us identify and quantify the PAH molecules. Additionally, we utilize UVIT images to assess the star formation associated with these PAH emitting regions. Our study focuses on three galaxies, namely NGC 628, NGC 1365, and NGC 7496, selected based on the availability of both JWST and UVIT images. Bright PAH emission regions are identified in the JWST images, and their corresponding UV emission is estimated using the UVIT images. We quantify the star formation properties of these PAH emitting regions using the UVIT images. Furthermore, we investigate the relationship between the star formation surface density ($\Sigma_{SFR}$) and the PAH ratios to better understand the impact of star formation on the properties of PAH molecules.}
   {Based on the resolved scale study on the PAH bright regions using JWST images, we found that the fraction of ionized PAH molecules is high in the star-forming regions with high $\Sigma_{SFR}$. We observed that emission from smaller PAH molecules is more in the star-forming regions with higher $\Sigma_{SFR}$. }
   {Our study suggests that the PAH molecules excited by the photons from SF regions with higher $\Sigma_{SFR}$ are dominantly smaller and ionized molecules. UV photons from the star-forming regions could be the reason for a higher fraction of the ionized PAHs. We suggest that the effect of high temperature in the star-forming regions and the formation of smaller PAH molecules in the star-forming regions might also be resulting in the higher fraction of emission in the F335$M_{PAH}$ band.}


   \maketitle
%
%------------------------------------------------------------------- 
\section{Introduction}
\label{Sect:Int}
Polycyclic Aromatic Hydrocarbon (PAH) molecules play a significant role in the infrared (IR) emission, contributing up to 20\% of the total IR emission in star-forming galaxies. This emission is characterized by a distinctive set of IR bands at 3.3, 6.2, 7.7, 8.6, 11.3 and 12.7 µm, accompanied by weaker IR features \citep{Madden2006,Smith2007,Li2020}. The formation of these IR emission features can be attributed to the stochastic heating of PAHs. The stretching modes of C-H bonds give rise to the 3.3, 8.6 and 11.3 $\mu$m PAH features, while the C-C bonds contributes to the 6.2 and 7.7 $\mu$m features \citep{Allamandola1989}.

There are different theories in place regarding the formation of PAH molecules. The widely accepted formation mechanism proposes carbon-rich asymptotic giant branch (AGB) stars as the cause for the formation of the PAH molecules. AGB stars possess the necessary conditions such as high collisional rates, cooling flows, shielding, and periodic shocks, to produce significant amounts of precursors (such as aliphatic hydrocarbons) to PAHs \citep{Latter1991}. Other formation mechanisms of PAH molecules include the fragmentation of graphitic grains in shocks \citep{Tielens1987, Jones1996}, accretion of $C^{+}$ in the diffuse interstellar medium  (ISM) \citep{Omont1986, Puget1989}, gas formation through ion-molecule reactions in dense clouds \citep{Herbst1991}, and the photoprocessing of interstellar dust mantles \citep{Greenberg2000}. However, most of these processes are not vetted properly due to the lack of dedicated observations.

The specific nature of each of the PAH features helps us to understand the properties of the PAH molecules such as their size and ionization. The population of larger, neutral PAH molecules are responsible for the 11.3 $\mu$m emission, while 7.7 $\mu$m is the resultant of vibrational bands of larger positively charged ions \citep{Galliano2008,Maragkoudakis2018,Draine2021,Rigopoulou2021, Maragkoudakis2022}. Similarly, most of the emission in the 3.3 $\mu$m band is attributed to small neutral PAHs in the range of 50-100 carbon atoms \citep{ricca2012,Draine2021}. \citet{ricca2012} found an inverse relation between the 3.3 $\mu$m/11.2 $\mu$m band ratio and PAH size suggesting that higher the band ratio, smaller will be the size of the PAH molecules. Hence the emission strength ratios of the PAH features can provide insights into the nature of PAH molecules.

The star-forming regions contribute significantly to the strength of the observed PAH emission features at 6.2, 7.7, 8.6 and 11.3 $\mu$m. Excitation of PAH molecules due to the ultraviolet (UV) photons from the massive young stars are considered as the driving force of PAH emission from the star-forming regions \citep{Tielens2008}. This association of PAH emission with the population of massive young stars is the underlying reason to use PAH features as tracers of star-formation in the galaxies \citep[e.g.][]{Peeters&Tielens2004, Xie2019}. However, the poor understanding of the nature and the properties of the PAH molecules remains a problem in using PAH emission as an indicator of star-formation activity. In this context, it is important to study the effect of star-formation on the properties of the associated PAH molecules. Studies have explored the impact of high-energy photons on the dust particles associated with HII regions that are linked to star-formation. \citep[e.g.][and references therein]{Riener2018}. The recent studies using the James Webb Space Telescope (JWST) also explored the properties of PAHs associated with the HII regions \citep[e.g.][]{Chastenet2023}. Although the emission observed in the UV range is considered as a direct indicator of star-formation \citep{Kennicutt1998}, previous studies investigating the connection between star-formation and PAH molecules were hindered by the spatial resolution limitations of telescopes available for observations in the UV and IR regimes.

JWST is revolutionizing our current understanding about the early Universe as well as the local Universe with its unprecedented spatial resolution in the IR regime \citep[e.g.][]{Harikane2023, Donnan2023, Finkelstein2023, Hoyer2023, Dale2023}. Among the four primary instruments in JWST, the Mid-Infrared Instrument \citep[MIRI;][]{Rieke2015}, Near-Infrared Camera \citep[NIRCam;][]{Rieke2005}, and Near-infrared Spectrograph \citep[NIRSpec;][]{Jakobsen2022} cover the IR spectral range where the PAH features are present. The wide range of filters provides us an opportunity to study the emission in different regions of the mid-IR regime. The coverage of filters, F335M ($\Delta\lambda$ $\sim$ 0.347 $\mu$m, FWHM $\sim$ 0.111"), F770W ($\Delta\lambda$ $\sim$ 1.95 $\mu$m, FWHM $\sim$ 0.269"), and F1130W ($\Delta\lambda$ $\sim$ 0.73 $\mu$m, FWHM $\sim$ 0.375") in JWST is designed to detect emission in the PAH features at 3.3, 7.7, and 11.3 $\mu$m, respectively. As we discussed earlier, these band ratios can be used to understand the nature of the PAH molecules. 

Regions undergoing active star-formation, characterized by the presence of young, hot, massive, and luminous O, B, A stars on the main sequence, emit a significant amount of UV radiation. As a result, these star-forming regions appear bright in UV images. Therefore, the ultraviolet continuum serves as a clear and direct tracer of recent star-formation in galaxies, typically within a time-frame of approximately 200 million years \citep{Kennicutt1998}. During the last decade, UltraViolet Imaging Telescope (UVIT), on board \textit{AstroSat}, opened up a new window of opportunity to perform high resolution UV studies of galaxies in the nearby Universe \citep[e.g.][]{Koshy2018a,Koshy2018b,Mondal2018,Yadav2021, Ujjwal2022, Prajwel2022, Devaraj2023, Rakhi2023, Robin2024}. Compared to its predecessor GALEX, UVIT offers $\sim$ 4 times better spatial resolution of $\sim$1.4". Due to its wide field of view (FOV = 28\arcmin), UVIT could provides the complete coverage of most of the nearby galaxies. 

In this study, we intend to associate the PAH band ratios with the star-formation properties to decipher how the star-formation affects the properties of the PAH molecules. In order to achieve this goal, we analyzed the high-resolution UV and IR images of three galaxies in the nearby Universe from UVIT and JWST space missions, respectively. A brief description about the sample is given in the Section \ref{sect:sample}. Data inventory is explained in Section \ref{sect:Data&analysis}, followed by the analysis \& results in Section \ref{sect:results}. Discussion and summary of the study are presented in Section \ref{sect:Discussion} and Section \ref{sect:summary}, respectively. A flat Universe cosmology is adopted throughout this paper with $H_{0}$ = 71 $km s^{-1} Mpc^{-1}$ and $\Omega_{M}$ = 0.27 \citep{Komatsu2011}.

\section{Sample}
\label{sect:sample}

To investigate the impact of star-formation on the characteristics of PAH molecules, it is essential to analyze high-resolution data in both the UV and IR ranges. Furthermore, to comprehend the relationship between star-formation and PAH molecules at a resolved level, the proximity of the galaxies plays a crucial role. Based on the availability of both UVIT and JWST observations, we have selected a sample of three nearby galaxies within a distance of 20 Mpc, i.e. NGC 628, NGC 1365, and NGC 7496, for this study. The comparable gas-phase metallicity of the sample of galaxies \citep[see][]{Groves2023} also enable us to discard the effect of metallicity on the PAH properties. A brief description of the galaxies used for this study is provided in the following subsections.

\begin{table*}[]
\centering
\begin{threeparttable}

\caption{Details of the sample of galaxies. The position information is taken from NASA/IPAC Extragalactic Database\tnote{1}. Distances are obtained from \citet{Anand2021}. Details regarding the JWST data for the sample of galaxies are detailed in \citet{Lee2023}.}
\label{tab:Table1}
\begin{tabular}{cccccccc}
\hline
\multirow{2}{*}{Name} & \multirow{2}{*}{RA (hh mm ss)} & \multirow{2}{*}{Dec (dd mm ss)} & \multirow{2}{*}{Distance~(Mpc)} & \multicolumn{3}{c}{Details of UVIT observation} \\
 &  &  &  & Obs. ID & Date of Obs. & Exp. Time (s) \\ \hline
NGC 628 & 01 36 41.747\tnote{1} & +15 47 01.18\tnote{1} & 9.8  & G06\_151 & 29-Nov-2016 & 1488.7 \\
NGC 1365 & 03 33 36.371\tnote{1} & --36 08 25.45\tnote{1} & 19.6  & G07\_057 & 31-Aug-2017 & 1162.6 \\
NGC 7496 & 23 09 47.290\tnote{1} & --43 25 40.58\tnote{1} & 18.7  & A07\_027 & 14-May-2020 & 3407.3 \\ \hline
\end{tabular}
\begin{tablenotes}
    \item[$^{1}$] \href{https://ned.ipac.caltech.edu/}{https://ned.ipac.caltech.edu/}
\end{tablenotes}
\end{threeparttable}

\end{table*}

\subsection {NGC 628}

NGC 628 is an Sc-type galaxy with an inclination of $9^{\circ}$, and at a redshift of z $\approx$ 0.00219, which corresponds to a distance of 9.8 Mpc \citep{Anand2021}. NGC 628 is the most prominent member of a small group of galaxies. The group is centered on NGC 628 and the peculiar spiral NGC 660. Two well-defined spiral arms observed in optical and UV images make NGC 628 a typical example of grand design spirals. NGC 628 has not gone through any recent interactions in the past 1 Gyr  \citep{Kamphuis1992}. \citet{Ujjwal2022} studied the star-forming regions in NGC 628 using UVIT FUV and NUV images and explored the secular evolution of NGC 628 driven by the spiral arms of the galaxy.

\begin{figure*}
\begin{center}
\includegraphics[width = 2\columnwidth]{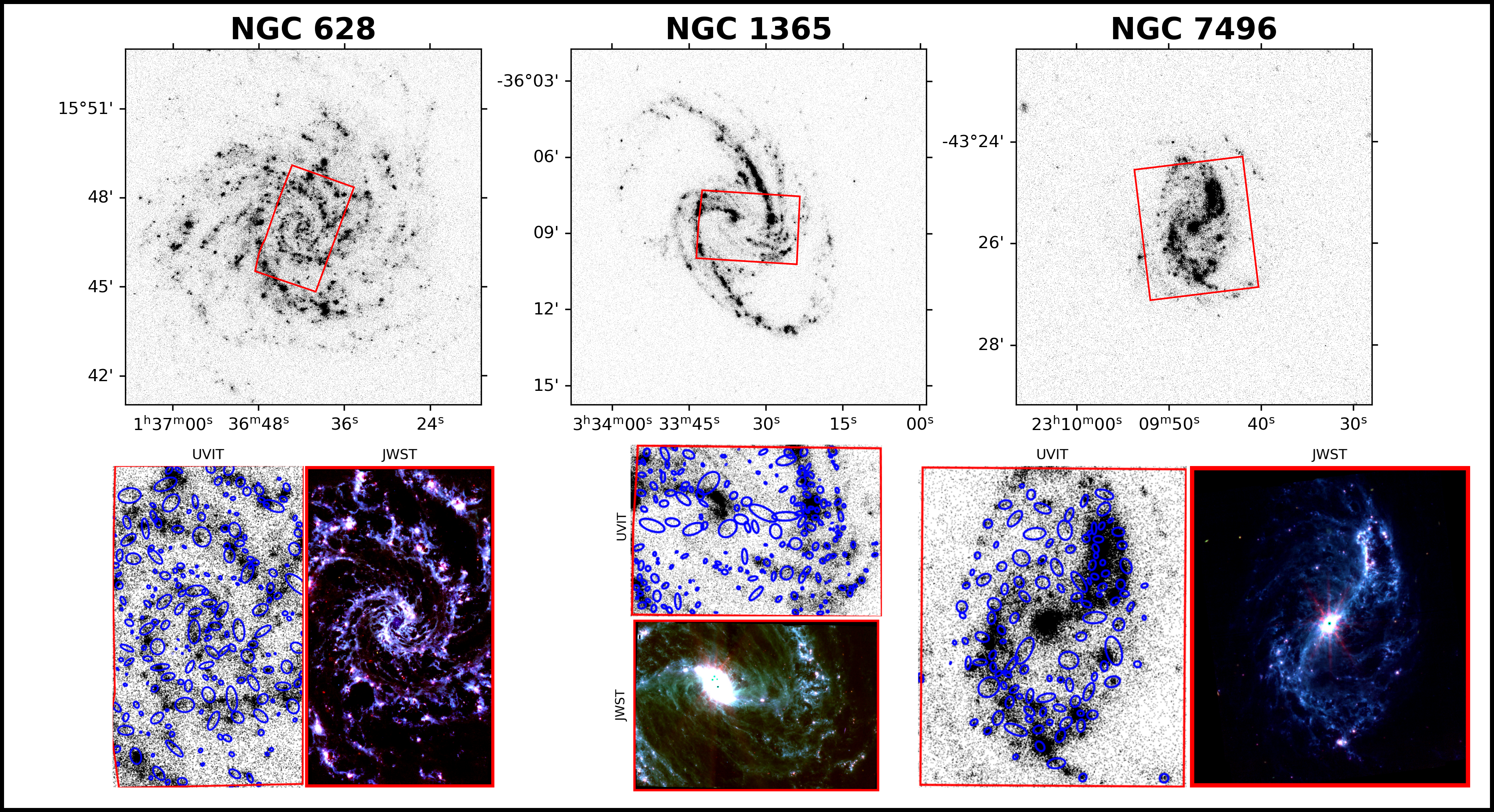}
\caption{ UVIT and JWST images of the sample of galaxies are shown in the top and bottom panels.  The field of JWST observation is marked in the UVIT FUV images. The UVIT footprint with JWST observation and the corresponding JWST color composite image of the sample of galaxies are shown in the bottom panel. The JWST color composite images are generated using F770W, F1130W, and F2100W filters, which are color-coded in blue, green, and red, respectively. Blue ellipses in the UVIT footprints in the lower panel represents the PAH bright regions identified using ProFound. For all the images North is up and East to the left.
}
\label{fig:color_comp}
\end{center}
\end{figure*}

\subsection{NGC 1365}
NGC 1365 is a member of the Fornax cluster \citep{Lindblad1999}. It is a nearby archetype of barred spiral SBb(s) Seyfert 1.8 galaxy at an approximate distance of 19.6 Mpc \citep{Anand2021}. Studies by \citet{Gao2021} shows that, when compared to the outer regions, NGC 1365 has an increased star-formation activity in the inner circumnuclear ring. \citet{Alonso2012} reported an intense star-formation activity in the inner parts of the galaxy using various IR star-formation indicators such as 6.2 and 11.3 $\mu$m PAH emission. They also observe that within the inner $\sim$ 5 kpc region of NGC 1365, the contribution of AGN to the IR emission is of the order of 5 percent.  

\subsection {NGC 7496}

NGC 7496 is a barred spiral SBb(s) Seyfert 2 galaxy at an estimated distance of 18.7 Mpc \citep{Anand2021}. It is reported to be a member of a group containing 9 galaxies \citep{Garcia1993}. The galaxy has a circumnuclear star-forming ring that possesses a high central gas surface density \citep{Sandstrom2023}. The galaxy also has a bar with a length of 1\arcmin ~\citep{Pence1984MNRAS.210..547P} that merges into the two spiral arms of the galaxy.  It has a fairly large central bulge and a prominent dust lane along the bar. The nuclear region exhibits emissions at 1300 MHz, solely due to the presence of an AGN. \citet{Singh2021} found that the star formation estimates of the central regions of the galaxy exhibit enhanced star-formation. They also suggest that there is a possibility of UV flux from AGN being interpreted as flux due to hot stars and hence the star formation.

\section{Data Inventory}
\label{sect:Data&analysis}

In this study, we utilize the MIRI F770W, F1130W, F2100W, and the NIRCAM F335M filters onboard JWST. The images of the sample of galaxies  are obtained from the the Canadian Astronomy Data Centre as part of the Physics at High Angular resolution in Nearby GalaxieS (PHANGS) archive  (PHANGS-JWST Treasury programme, PI: J C Lee). The details pertaining to the reduction of JWST images are explained in \citet{Lee2023}. The images obtained from the PHANGS-JWST Treasury program have an angular resolution of 0.111", 0.25", and 0.36" for F335M, F770W, and F1130W bands, respectively \citep{Lee2023}. The filters F770W and F1130W are primarily designed to sample the sources with PAH emission at 7.7 and 11.3 $\mu$m, respectively \citep[see][]{Chastenet2023}. In order to extract the PAH features from the F335M imaging, starlight continuum needs to be removed from the observed data. Based on the NIRCam F335M (measuring the main PAH emission feature) along with the F300M and F360M filters, \citet{Sandstrom2023} devised a new method to remove starlight continuum to obtain the F335M$_{PAH}$ images. We made use of the prescription by \citet{Sandstrom2023} to remove the contamination of starlight from the F335M images and is given below,

\begin{center}
\begin{equation}
F335M_{PAH} = F335M - F335M_{cont}
\end{equation}
\begin{equation}
F335M_{cont} = A \times F300M + B \times F360M
\end{equation}
\end{center}

The coefficients provided by \citet{Sandstrom2023} are A = 0.35 and
B = 0.65. F335M$_{PAH}$ and F335M$_{cont}$ maps are in units of MJy sr$^{-1}$.

We use UVIT data of the sample of galaxies NGC 628 (PI: ASK pati), NGC 1365 (PI: C S Stalin) and NGC 7496 (PI: Erik Rosolowsky) obtained from the \textit{AstroSat} ISSDC archive \citep{Kumar2012SPIE.8443E..1NK}. We made use of F148W ($\lambda_{peak} \sim 1481 \textup{~\AA} $, $\Delta\lambda \sim 500 \textup{~\AA}$) images for the study. The Level 1 data was reduced using the software CCDLab - version 17 \citep{Postma2017}. The VIS images have been utilized to incorporate drift correction. Each image is flat fielded, followed by distortion correction and pattern noise using the calibration files in CCDLab \citep{Girish2017, Postma2017,Tandon2017,Tandon2020}. Astrometric solutions were also made using CCDLab by making use of the Gaia DR3 detections in the observed field \citep{Postma2020}. The resolution of UVIT FUV images of 1.4'' enable us to resolve the star-forming regions upto $\sim 66$ pc, 127 pc, and 132 pc in NGC 628, NGC 7496, and NGC 1365, respectively. The details pertaining to the UVIT observations, along with the position and distance information of the sample of galaxies are listed in Table \ref{tab:Table1}.  Figure \ref{fig:color_comp} represents the UVIT and JWST images of the sample of galaxies. The top panel of Figure \ref{fig:color_comp} represents the UVIT FUV images, which are marked with the footprints of JWST observations. JWST color composite images of the sample of galaxies were generated using F770W, F1130W, and F2100W filters and are shown in the bottom panel of Figure  \ref{fig:color_comp}.

To match the resolution of the images used in this study, JWST images are convolved to the point spread function (PSF) of the UVIT FUV filter with a FWHM of 13 and 22 pixels for MIRI and NIRCam images respectively. Among the JWST images of the three sample of galaxies, NGC 7496 is the only target having an observed area without any considerable source contamination. We estimated a background value of $\sim$ 0.07 MJy$sr^{-1}$ in the F770W image of NGC 7496. We used this estimate to remove the background from all the MIRI images of our sample of galaxies. In the final image, only pixels with signal-to-noise (S/N) >= 3 are considered for further analysis. In the case of NIRCam F335M image, we estimated a background of $\sim$ 0.014 $MJy sr^{-1}$ in the image of NGC 7496. Similar to the MIRI images, we used this background estimate to select the pixels with  S/N $\geqq$ 3 in the NIRCam image of all three galaxies. To remove any possible contamination due to the Active Galactic Nuclei (AGN) present in the galaxies NGC 1365 and NGC 7496, the central 15" region of the images has been masked.

\section{Analysis \& Results}
\label{sect:results}

\subsection{Tracing the PAH emission using JWST images}
\label{subsect:Identification_PAH}

UV photons are significantly absorbed by PAH molecules. The excited PAH molecules de-excites through vibrational transitions, resulting in the production of molecular emission features in the IR \citep{2007Draine_li}. To correlate the properties of the PAH molecules with the corresponding UV emission, it is important to identify the bright regions in the JWST images. Identifying the star-forming regions using UVIT images and then correlating them with the corresponding emission in JWST images will limit us only to the PAH emission associated with star-formation. We identified regions with the brightest PAH emission from JWST images and then estimated the UV emission associated with the same region from the corresponding UVIT images.

To extract the brightest regions from the MIRI 770W images of the sample of galaxies, we used the ProFound package. ProFound is an astronomical data processing tool available in the R programming language \citep{Robotham2018}. ProFound locates the image's peak flux areas and identifies bright emission regions as individual source segments. The total photometry is then estimated using iterative expansion of the observed segments. We used the subroutine of the same name in ProFound to identify the regions with bright PAH emission in the F770W image. We defined the criterion that the identified regions should cover at least 24 pixels. This criterion was selected in order to account for the resolution of FUV filters. The main emphasis in this approach is to determine the fewest number of pixels necessary to enclose a circle with a diameter that corresponds to the resolution of the FUV filter. The PAH segments identified using the F770W image is overlaid on the F1130W and $F335M_{PAH}$ images and the corresponding PAH emission is quantified.

\subsection{UV emission from the PAH bright regions}
\label{subsect:Identification}

We overlaid the segments obtained for the MIRI F770W images on the UVIT FUV images, and the basic information of the identified regions such as position, flux, and extent are obtained \citep[see][]{Ujjwal2022}. The segments overlaid on the UVIT images are represented in the lower panel of Figure \ref{fig:color_comp}. After dilating to obtain a total flux measurement, the output of the analysis performed with ProFound gives us the number of pixels contained in the identified segments. We made use of the total number of pixels to estimate the area of the identified star-forming regions. Then, the magnitude of each identified region has been estimated.

Emission in the UV regime is strongly affected by extinction. In order to estimate star-formation rate (SFR) in the star-forming regions using UV emission, it is mandatory to account for both internal extinction and the line of sight extinction in the Milky Way. The $F_{H{\alpha}}/F_{H{\beta}}$ for the identified star-forming regions are compiled from \citet{Santoro2022}. An intrinsic Balmer ratio of 2.86 for case B recombination at an electron temperature $T_{e}$ = 10,000 K and density $n_{e}$ = 100  $cm^{-3}$ is considered for the internal extinction correction \citep{Osterbrock1989}. Extinction in $H_{\alpha}$ is estimated as follows,

\begin{center}
\begin{equation}
    A(H\alpha) = \frac{K_{H\alpha}}{-0.4 \times (K_{H\alpha}-K_{H\beta})} \times log\frac{F_{H\alpha}/F_{H_\beta}}{2.86}
\end{equation}
\end{center}

where $K_{H\alpha}$ = 2.53 and $K_{H\beta}$ = 3.61 are the extinction coefficients for the Galactic extinction curve from \citet{Cardelli1989}.The $A(H\alpha)$ is then converted to $A_{FUV}$ using the coefficients obtained from \citet{Cardelli1989}. The line of sight extinction has also been accounted using \citet{Cardelli1989} with the $A_{v}$ values compiled for each galaxy from \citet{Schlafly2011}. Hereafter, we use the extinction corrected magnitudes for further analysis. The SFR in each region is estimated using the relation obtained from \citet{Karachentsev2013} and is given below.

\begin{equation}
log(SFR_{FUV}(M_{\odot}yr^{-1})) = 2.78-0.4mag_{FUV}+2log(D)
\end{equation}

where $mag_{FUV}$ denotes the extinction corrected magnitude and D is the distance to the galaxy in Mpc. The Profound package provides the total number of pixels containing 100\% of the flux for each of the identified regions. Subsequently, we have calculated the area by considering the plate scale of the UVIT images. The star-formation surface density ($\Sigma_{SFR}$) of the identified regions were determined by dividing the estimated SFR by the corresponding area of each region (in {\normalfont kpc$^2$}).

\begin{figure*}
\begin{center}
\includegraphics[width = 2\columnwidth]{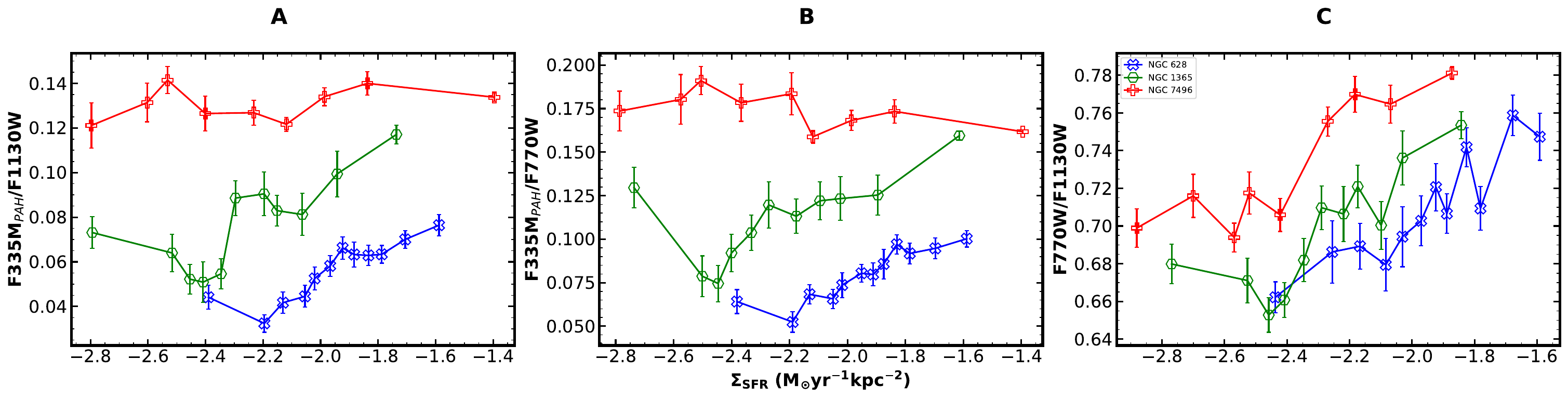}
\caption{Figure represents the variation of the running median of the band ratios corresponding to the $\Sigma_{SFR}$. A, B and C panels represents the band ratios $F335M_{PAH}$/F1130W, $F335M_{PAH}$/F770W, and F770W/F1130W, respectively. Blue x, green hexagon, and red +, symbols represent the respective galaxies NGC 628, NGC1365 and NGC 7496. The associated standard mean error is also represented. It is clear that with the increase of $\Sigma_{SFR}$, the band ratios increases in all three galaxies.}
\label{fig:running_med}
\end{center}
\end{figure*}

\subsection{Band ratios and the UV emission}

We have discussed how the PAH band ratios can be utilized to understand the properties of the contributing PAH molecules in Section \ref{Sect:Int}. For example, in a region of interest, a higher fraction of ionized PAH is indicated by a higher value of F770W/F1130W. Similarly, the band ratios F335M$_{PAH}$/F770W and F335M$_{PAH}$/F1130W can be used to identify the size of the dominant PAH molecules in a region of interest, i.e., higher value of F335M$_{PAH}$/F770W or F335M$_{PAH}$/F1130W suggest that smaller neutral PAH molecules are dominant compared to larger PAH molecules \citep[e.g.][]{Micelotta2010}.

To understand how the nature of the PAH molecules relates to the star-formation environment of the sample of galaxies, we plotted the $\Sigma_{SFR}$ versus the running median of the band ratios as shown in Figure \ref{fig:running_med}. In our analysis, we focus solely on the regions within the $16^{th}$ - $84^{th}$ percentile range of the total distribution. This approach helps us to address any potential offset that may arise from the extreme values observed in the JWST bands. The bin selection has been made in such a way that the number of data points in each bin remains the same. Figure \ref{fig:running_med}A represents the variation of the running median of the PAH band ratios F335M$_{PAH}$/F1130W. It is observed that the band ratios are higher for regions with high $\Sigma_{SFR}$ for the sample of galaxies. It is also evident in Figure \ref{fig:running_med}A that there exist an offset between the band ratios of all three galaxies. An increase in the band ratio can be correlated to the decrease in the size of the PAH molecules. We observe that the band ratio $F335M_{PAH}$/F1130W increases in the order NGC 628  $<$  NGC 1365 $<$ NGC 7496. It implies that NGC 7496 hosts smallest PAH molecules, where as NGC 628 contains the largest PAH molecules in the sample of galaxies used in this study. A similar trend is observed in F335M$_{PAH}$/F770W (Figure \ref{fig:running_med}B). It may be noted that the variation of the band ratio F335M$_{PAH}$/F770W is relatively less in NGC 7496, when compared to NGC 628 and NGC 1365.

From Figure \ref{fig:running_med}C, it is evident that the ratio F770W/F1130W increases with an increase in the $\Sigma_{SFR}$. This in-turn suggests that the fraction of ionized PAH molecules are higher in the star-forming regions with higher $\Sigma_{SFR}$. The high energetic UV photons from the star-forming regions could be the reason for the observed high ionization in the region. Also, we observe a shift in the estimates of the band ratios between the galaxies. It suggests that NGC 628 hosts more neutral PAH molecules whereas NGC 7496 contains least amount of neutral PAH molecules. This finding can be linked to the previous result indicating less variability in the F335M$_{PAH}$/F770W ratio for NGC 7496. One possible explanation for this observation could be the influence of an active nucleus within the galaxy \citep[e.g.,][]{Garcia1_2022}.

\begin{figure*}
\begin{center}
\includegraphics[width = 1.5\columnwidth]{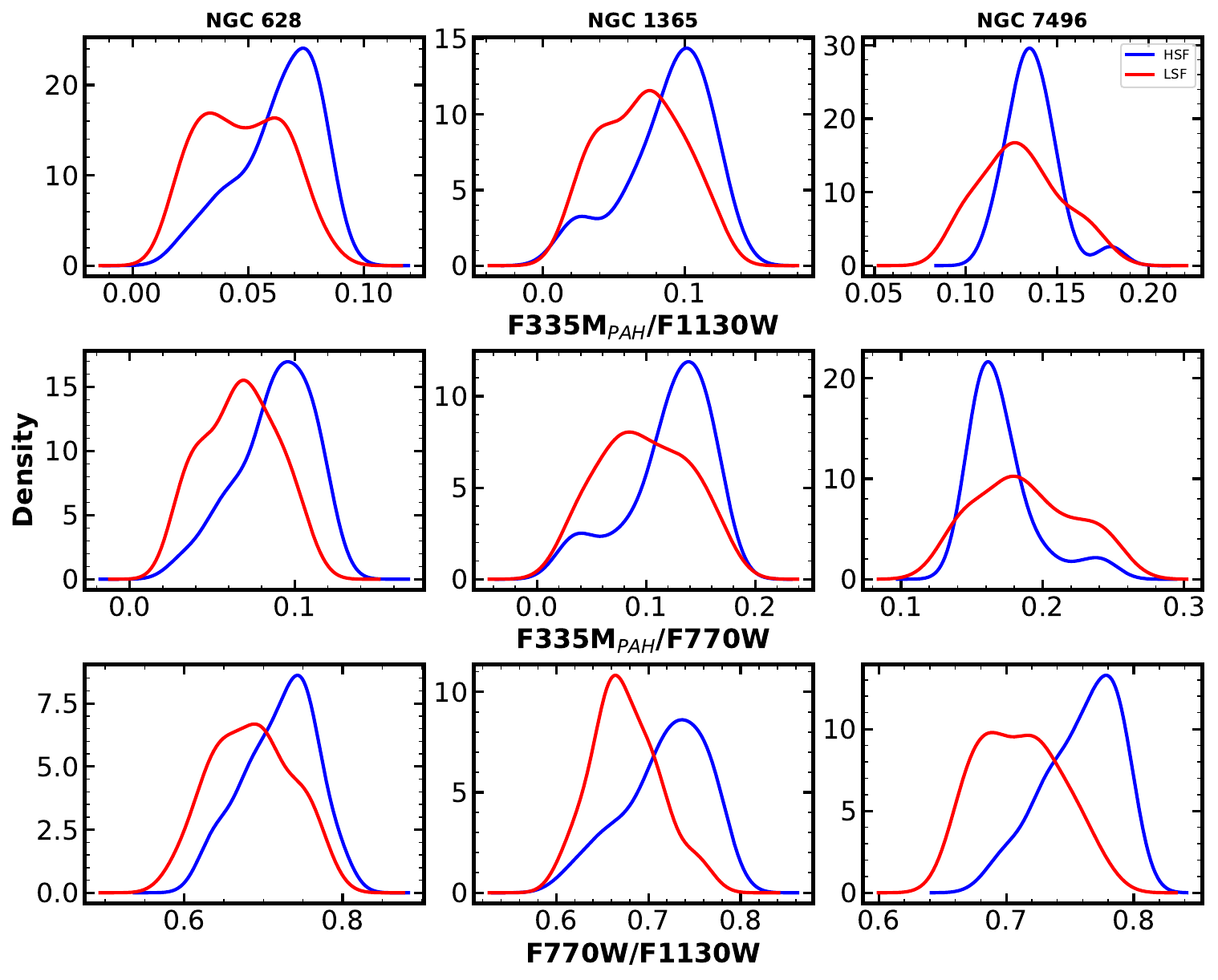}
\caption{ KDE plot representing the band ratios in the sample of galaxies with respect to the $\Sigma_{SFR}$. The PAH bright regions are classified into high star-forming and low star-forming regions, based on the $\Sigma_{SFR}$ of the associated star-forming regions. Blue represents the regions with high $\Sigma_{SFR}$ whereas red represents the regions with low $\Sigma_{SFR}$.}
\label{fig:Pah_rat}
\end{center}
\end{figure*}

Results from the Figure \ref{fig:running_med} suggests a difference in the population of PAH molecules with respect to the $\Sigma_{SFR}$ of the associated star-forming regions. To investigate the influence of the star-forming environment on the properties of PAH molecules, we divided the star-forming regions within the sample of galaxies into two categories: high star-forming (HSF) regions and low star-forming (LSF) regions. This categorization was based on the median $\Sigma_{SFR}$ observed in each galaxy's star-forming regions. Star-forming regions with $\Sigma_{SFR}$ values lower than the median $\Sigma_{SFR}$ of the respective galaxy were classified as LSF regions, while regions with equal to or higher $\Sigma_{SFR}$ values were considered HSF regions. The median $\Sigma_{SFR}$ values for NGC 628, NGC 1365, and NGC 7496 were found to be 0.011, 0.005, and 0.004 $M_{\odot} yr^{-1} kpc^{-2}$, respectively.

First two panels of Figure \ref{fig:Pah_rat} depict the Kernel Distribution Estimation (KDE) distribution of band ratios $F335M_{PAH}$/F1130W, and $F335M_{PAH}$/F770W for star-forming regions classified as HSF and LSF within the sample of galaxies. From Figure \ref{fig:Pah_rat} it is evident that the band ratios $F335M_{PAH}$/F1130W, and $F335M_{PAH}$/F770W exhibit higher values in star-forming regions belonging to HSF compared to those belonging to LSF, suggesting that the smaller PAH molecules are dominant in the star-forming regions with higher $\Sigma_{SFR}$. In this analysis also, $F335M_{PAH}$/F770W ratio of NGC 7496 show a deviation from the general trend. As we discussed earlier, the higher fraction of the ionized PAH molecules might be the reason for this deviation. In order to check whether the band ratios of PAH bright regions belonging to HSF and LSF constituted by two different population of PAH molecules, we performed a two-sample Kolmogorov–Smirnov test (K-S test) on the properties of the star-forming regions. The probability predicted by K-S test for the band ratios $F335M_{PAH}$/F1130W and $F335M_{PAH}$/F770W of all three sample of galaxies are < 10\% \& < $10^{-1}$\%, respectively. These results suggest that the population of PAH molecules linked to star-forming regions with higher $\Sigma_{SFR}$ differs from those associated with lower $\Sigma_{SFR}$. It could be interpreted as the size of the PAH molecules being different in high and low star-forming environments.

Lower panel of Figure \ref{fig:Pah_rat} represents the Kernel density estimate (KDE) distribution of the band ratio F770W/F1130W. It is observed that the star-forming regions with higher $\Sigma_{SFR}$ exhibits higher values for F770W/F1130W in all three sample of galaxies compared to the regions with lower $\Sigma_{SFR}$. In order to check whether the band ratios of PAH bright regions belonging to HSF and LSF constituted by two different population of PAH molecules, we performed a two-sample K-S test on the properties of the star-forming regions. For all three galaxies, K-S test performed on the F770W/F1130W band ratio of HSF and LSF, resulted a probability of < $10^{-1}$ \%. This in turn suggests that the population associated with star-forming regions having high $\Sigma_{SFR}$ are distinct when compared to those associated with star-forming regions with relatively lesser $\Sigma_{SFR}$. The difference in the fraction of ionized molecules could be the reason for the observed distinction.

\section{Discussion}
\label{sect:Discussion}

In this study, we analyze the impact of star-formation on the resolved scale properties of polycyclic aromatic hydrocarbon (PAH) molecules in a selected sample of three nearby galaxies. Specifically, we investigate how star-formation influences the properties of PAH molecules that are spatially associated with the regions undergoing active star-formation. Based on the availability of both JWST and UVIT images, a sample of three  nearby galaxies - NGC 628, NGC 1365, and NGC 7496 have been selected for the study. Even though the JWST images offer higher resolution compared to UVIT images, the JWST images are degraded to the resolution of UVIT images for the analysis. PAH bright regions are identified from the JWST images and the corresponding star-formation rate is estimated using the UVIT images. We examine the correlation between the $\Sigma_{SFR}$ and the PAH band ratios to gain insights into the influence of star-formation on the ionization and the size of PAH molecules. 

Even though the F770W and the F1130W bands predominantly trace the 7.7 and 11.3 $\mu$m PAH emission, it is important to acknowledge the potential presence of additional contributions arising from hot dust and/or continuum emission. Utilizing the galaxy sample extracted from the SINGS survey, Whitcomb et al. (2023, in preparation) employed the PAHFIT method \citep{Smith2007} to extract the contribution of hot dust and/or continuum emissions originating from star-forming regions to the spectral bands of JWST. They found that approximately 80\% of the F770W emissions are predominantly attributed to PAHs, while roughly 70\% of the emissions observed in F1130W are due to PAH molecules. This result aligns with emission-line measurements of \citet{GarciaBernete2022}, who employed PAHFIT for analyzing the Spitzer/IRS spectra for NGC 1365 and NGC 7496. In the context of JWST data, more robust estimates on the contamination will be a subject of future studies. To assess the potential impact of contamination on our findings, we considered potential contamination rates of 20\% and 30\% attributed to dust continuum and line emission fraction in the F770W and F1130W filters, respectively. We found that even-though the specific values of flux ratios varies, the observed trend in the corresponding PAH band ratios remains unaltered by contamination.

Furthermore, beyond the issue of continuum contamination, there exists the possibility of silicate absorption at 9.7 $\mu$m affecting the F1130W flux. Smith et al. (2007) have suggested that this absorption phenomenon is likely to exert only a limited influence on the majority of star-forming galaxies. Furthermore, recent findings by \citet{Groves2023} indicate that H II regions in the sample of galaxies used in this study are characterized by a median E(B-V) value in the range of 0.2-0.3. This signifies a relatively low level of dust attenuation. Hence, we suggest that the silicate absorption feature might not be significantly affecting the results presented in this study. However, more extensive imaging and spectral studies are required to estimate the effect of 9.7 silicate absorption feature contributing to the filter F1130W.

\citet{Chastenet2023} studied the PAH molecules in the HII regions using PHANGS-JWST images for the sample galaxies in our study. They found that the areas with harder radiation fields near HII regions had a higher proportion of smaller PAH molecules, based on the ratios of $F335M_{PAH}$/F1130W and $F335M_{PAH}$/F770W. The results from our study also suggest that the smaller PAH molecules are dominant in the star forming regions with high $\Sigma_{SFR}$. However, the apparent increase in the proportion of smaller PAHs within harsh environments contradicts theoretical understanding that the smaller PAHs typically exhibit shorter lifespans in the hot gas environment \citep[e.g.][]{Micelotta2010}. In this study, while considering the PAH band ratios, the effect of temperature has not been taken into account. An increase in the band ratio can also be due to the decrease in size of PAHs or increase in the temperature of the region. In this context, it is important to discuss about the temperature of the regions of our interest in this analysis. In the star-forming regions with higher $\Sigma_{SFR}$, the PAH molecules could be hotter due to the increase in the average energy of the photons. This increase in temperature could result in a higher emission at lower wavelengths such as 3.3 $\mu$m \citep[e.g.][]{Draine2021}. But, we cannot confirm that the temperature as a possible factor for the observed higher fraction of smaller PAH molecules in the harsh environments. To account for the drawbacks of PAH formation, \citet{Jones1996} proposed a mechanism for the formation of PAHs involving grain–grain collisions and shattering. This concept, also discussed by \citet{Guillet2011}, suggests that in interstellar shocks, large grains undergo fragmentation, leading to the generation of numerous small grains through the shattering process. Formation of smaller PAH molecules through fragmentation can also be a possible reason for the observed higher fraction of smaller PAH molecules in the regions with high $\Sigma_{SFR}$. The enhanced detection of smaller PAH molecules support the observations by \citet{Maragkoudakis2023}, that the smaller PAHs have higher impact on the PAH band strengths.

In this context, we have to keep in mind that PAH could be affected by different factors. Metallicity is one such factor \citep[e.g.,][]{Draine2007r}. Metal poor galaxies tend to show very weak or no PAH features in their IR spectra \citep[e.g.][]{Wu2006, Hao2009}. \citet{Groves2023}, estimated the metallicity (12+log(O/H)) for the PHANGS sample of galaxies  based on S-calibration. For NGC 628, NGC 1365, and NGC 7496 they reported 12+log(O/H) values of 8.478, 8.477, and 8.507, respectively. Hence, metallicity might not be playing a significant role in the results presented in this study. The presence of an active galactic nuclei (AGN) is yet another factor which could influence the existence of PAH molecules. Since, two among the three galaxies used in this study host AGNs, it needs to be noted that the UV and soft X-ray photons in the strident environment around AGNs could destroy PAHs \citep{Voit1992, Siebenmorgen2004}. Studies also found PAH emission in the close proximity of AGNs within  $\lesssim$ 10 pc of the nucleus \citep{Esquej2014, Jensen2017}. This is suggestive that PAHs could survive in the near vicinity of AGNs and could be excited by the photons from them. 

This study connects the high resolution JWST images with the high resolution UV images of the nearby galaxies. In this study, we couldn't make use of the full resolution capabilities of JWST as we gave importance to the comparison using UVIT images. Yet, the unprecedented resolution provided by the JWST and UVIT enabled us to understand the effect of star-formation in the properties of the associated PAH molecules. Future works needs to be made to decipher how the PAH band ratios are interconnected with the properties of the stellar population such as age, and metallicity. More works are in the offing to decipher the role of age and metallicity of the stellar population in addressing the abundance of PAH molecules in galaxies.

\section{Summary}
\label{sect:summary}
A summary of the main results obtained from our study is given below

\begin{itemize}
    
\item Based on the availability of JWST and UVIT images, we selected the galaxies NGC 628, NGC 1365, and NGC 7496 to study the effect of star-formation with the associated PAH emission.

\item Bright PAH emission regions have been identified using the JWST F770W images and the corresponding UV emission has been quantified using UVIT FUV images.

\item Based on the correlation between $\Sigma_{SFR}$ and the corresponding PAH band ratios, we found that the fraction of ionized PAH molecules is higher in the star-forming regions with high star-formation rate density compared to those with lower $\Sigma_{SFR}$.

\item In the star-forming regions with higher $\Sigma_{SFR}$, we have observed increased emission from smaller PAH molecules. This could be influenced by the elevated temperatures within the star-forming regions or the formation of smaller PAH molecules.
 
\end{itemize}

\begin{acknowledgements}
  We thank the anonymous referee for the valuable comments that improved the scientific content of the paper. UK acknowledge the help and support from Adam Leroy and Erik Rosolowsky for their time and effort for sharing the PHANGS JWST images used in this study. The authors thank Koshy George, Savithri H Ezhikode and Arun Roy for their suggestions throughout the course of this work.  UK acknowledges the Department of Science and Technology (DST) for the INSPIRE FELLOWSHIP (IF180855). SSK and AKR acknowledge the financial support from CHRIST (Deemed to be University, Bangalore) through the SEED money project (No: SMSS-2220, 12/2022). SSK and RT, acknowledge the financial support from Indian Space Research Organisation (ISRO) under the \textit{AstroSat} archival data utilization program (No. DS-2B-13013(2)/6/2019). This publication uses the data from the UVIT, which is part of the\textit{AstroSat} mission of the ISRO, archived at the Indian Space Science Data Centre (ISSDC). We gratefully thank all the individuals involved in the various teams for providing their support to the project from the early stages of the design to launch and observations with it in the orbit. This research has used the NASA/IPAC Extragalactic Database (NED), funded by the National Aeronautics and Space Administration and operated by the California Institute of Technology. We thank the Center for Research, CHRIST (Deemed to be university) for all their support during the course of this work.
  
\end{acknowledgements}

\bibliographystyle{aa}
\bibliography{bibtex} % if your bibtex file is called example.bib

\end{document}